\def\year{2020}\relax
\documentclass[letterpaper]{article} 
\usepackage{aaai20}  
\usepackage{times}  
\usepackage{helvet} 
\usepackage{courier}  
\usepackage{color}
\usepackage[hyphens]{url}  
\usepackage{graphicx} 
\usepackage{amsmath,bm}
\usepackage{url}
\usepackage{comment}
\usepackage{amsfonts}
\usepackage{todonotes}
\usepackage{hyperref}
\def\UrlFont{\rm}  
\usepackage{graphicx}  
\newcommand{\colorann}[3]{\textcolor{#1}{${}^{#2}[$#3$]$}}
\newcommand{\kz}[1]{\colorann{red}{kz}{#1}}

\nocopyright

\title{Challenges in Combating COVID-19 Infodemic \\ - Data, Tools, and Ethics}
\author{Kaize Ding,
Kai Shu,
Yichuan Li,
Amrita Bhattacharjee, and 
Huan Liu \\
Computer Science and Engineering, Arizona State University, USA\\
\{kding9, kai.shu, yichuan1, abhatt43, huanliu\}@asu.edu
}

\begin{document}

\maketitle

\begin{abstract}
While the COVID-19 pandemic continues its global devastation, numerous accompanying challenges emerge. One important challenge we face is to efficiently and effectively use recently gathered data and find computational tools to combat the COVID-19 infodemic, a typical information overloading problem. Novel coronavirus presents many questions without ready answers; its uncertainty and our eagerness in search of solutions offer a fertile environment for infodemic. It is thus necessary to combat the infodemic and make a concerted effort to confront COVID-19 and mitigate its negative impact in all walks of life when saving lives and maintaining normal orders during trying times. In this position paper of combating the COVID-19 infodemic, we illustrate its need by providing real-world examples of rampant conspiracy theories, misinformation, and various types of scams that take advantage of human kindness, fear, and ignorance. We present three key challenges in this fight against the COVID-19 infodemic where researchers and practitioners instinctively want to contribute and help. We demonstrate that these three challenges can and will be effectively addressed by collective wisdom, crowd sourcing, and collaborative research. 
\end{abstract}

\section{Introduction}
Coronavirus disease 2019 (COVID-19) is an infectious disease caused by severe acute respiratory syndrome coronavirus 2 (SARS-CoV-2). The World Health Organization (WHO) recently declared the COVID-19 outbreak a Public Health Emergency of International Concern (PHEIC) and a pandemic due to its high morbidity and mortality rates. As of April 15, 2020, more than 2.04 million cases have been reported across 210 countries and territories, resulting in over 133,000 deaths\footnote{https://en.wikipedia.org/wiki/Coronavirus\_disease\_2019}. These numbers are continuing to rise and the health systems in many countries are overwhelmed to provide treatment. Concomitant with the pandemic are many unknowns that create a conducive environment for misinformation, fake news, political disinformation campaigns, scams, etc. Those malicious contents instigate fears or anger, capitalize on human vulnerability, and exploit human emotion, kindness, and/or wishes for miracles. 

As the coronavirus spreads like fire in the world, disinformation machines also accelerate their campaigns on various fronts, rendering a new infodemic battlefield. Social media platforms such as Facebook/Instagram, Twitter, and Google/YouTube have been abused to disseminate erroneous contents. When the whole world is scrambling to fight the COVID-19 pandemic, governments and WHO also have to combat an infodemic, which is defined as ``an overabundance of information — some accurate and some not—that makes it hard for people to find trustworthy sources and reliable guidance when they need it''~\cite{donovan2020infodemic}. The COVID-19 infodemic causes confusion, sows division, incites hatred, promotes unproven cures, and provokes social panic, which directly impacts emergency response, treatment, recovery, and financial and mental health during the difficult time of self-isolation. Therefore, combating the COVID-19 infodemic is a challenging yet imperative task.


In this paper, we first present some COVID-19 related examples to illustrate the variety and range of infodemic cases in representative categories: conspiracy theories and misinformation, and scams and security attacks to reinforce the urgency and need for addressing the COVID-19 infodemic via scalable and timely solutions. We then discuss the essential challenges in designing and developing corresponding AI solutions from three perspectives: data, computational tools, and ethics. The last challenge of ethics is particularly easy to overlook when we rush to confront the immediate threats. Therefore, it is important to understand unintended consequences when developing AI solutions to ensure sustainable and healthy use and deployment. Last, we use some current efforts to demonstrate the feasibility of addressing the three challenges in combating the COVID-19 infodemic; by understanding the challenges and what we have, we also appreciate the importance of collaborative research to effectively and efficiently combat the COVID-19 infodemic.

\section{Examples of COVID-19 Infodemic}

To illustrate what the COVID-19 infodemic looks like, how expansive, active, and devastating it is, and why it is important to thwart or mitigate its present threats, we first present various examples regarding conspiracy theories and misinformation, and scam and security attacks.  


\subsection{Conspiracy theories and misinformation}


With the spread of COVID-19 pandemic,
the World Health Organization (WHO) recently warned of an ``infodemic'' of rampant conspiracy theories about the coronavirus. Those conspiracy theories have appeared in both social media and mainstream news outlets and are often intertwined with geopolitics. One example is about how the new coronavirus originated: according to a Pew Research Center survey, nearly three-in-ten Americans believe COVID-19 was a bio-weapon made in the lab. Some top 10 conspiracy theories include SARS-CoV-2 virus was created as a biologic weapon from a lab, GMOs are the culprit, COVID-19 actually doesn't exist, and coronavirus is a plot by big Pharma~\cite{lynas2020covid}.







Coronavirus misinformation is also flooding the internet through social media, text messages, and propagated by celebrities, politicians, or other prominent public figures. According to the report in~\cite{Kornbluh-Goodman2020}, ``among outlets that repeatedly share false content, eight of the top 10 most engaged-with sites are running coronavirus stories." For instance, there are plenty of supposed ``cures'' on social media that will likely mislead people to risk their lives for quick fixes. Disregarding the National Institutes of Health (NIH) warning of many hearsay cures without evidence of curing being effective, there are endless claims such as herbs and teas, or something of the sort that can prevent the coronavirus. Recently, some wireless towers were damaged in the UK due to a false claim that radio waves sent by 5G technology are causing small changes to people's bodies that make them succumb to the virus.

\subsection{Scam, spam, phishing, and malware attacks}




As more and more people start working or studying from home, cyber criminals recently shift focus to target remote workers. Different attacks such as scam, spam, phishing and malware, which prey on people's willingness to help, fear of supply shortage, and moments of weakness, have become increasingly active. Researchers have found that the volume of coronavirus email scams nearly tripled in one week, with almost 3\% of all global spam now estimated to be COVID-19 related. During the coronavirus pandemic, as state governments and hospitals have scrambled to obtain masks and other medical supplies, scammers attempted to sell a fake stockpile of 39 million masks to a California labor union. 
According to The Hill~\cite{miller2020virtual}
, ``Hackers are taking advantage of the increased reliance on networks to target critical organizations such as health care groups and members of the public, stealing and profiting off sensitive information and putting lives at risk." 

\section{Data, Tool, and Ethics Challenges}
The scale, volume, and reach of the COVID-19 infodemic entails the reliance on AI and machine learning (ML) algorithms to react promptly and respond rapidly. The success of AI and ML algorithms requires large amounts of multi-modal data for their efficiency and effectiveness, which  introduces \textit{a data challenge}. Data extraction and curation from multi-source data needs different computational tools to accurately categorize and sort out various types of data, which presents \textit{a tool challenge}. When we rush to deal with present threats, we should be aware of potential side-effects, unexpected consequences, and biases of our solutions, which suggests \textit{an ethic challenge}. In this section, we will discuss these three challenges in detail. 

\subsection{Data challenge}



Though numerous COVID-19 data sources are available online, their datasets are available on various websites for different needs. 
The major data challenge of  isolated data sources is the awareness of their existence. Another related issue is that they are collected from different sources or under different crawl settings. For example, Allen Institute for AI (AI2) released the scholarly articles dataset\footnote{https://allenai.org/data/cord-19} collected from PMC, medRxiv and bioRxiv; the Frontiers\footnote{https://coronavirus.frontiersin.org/} provided the latest research articles, and LitCovid~\cite{RN12503} collected the scientific information from PubMed. 
Combining different data sources leads to higher quality of data and better coverage. 


To address the data challenges, we need to overcome some shortcomings: \textit{ disorganization} -- most of them merely list all the collected datasets on their websites without information summarizing the relationships among them; \textit{specificity} -- data collected for a specific topic,  
for example, Amazon provides the epidemic dataset on  cloud\footnote{https://aws.amazon.com/blogs/big-data/a-public-data-lake-for-analysis-of-covid-19-data/} and COVID-19 GIS Hub\footnote{https://coronavirus-disasterresponse.hub.arcgis.com/} only contain the academic findings and geospatial-related datasets respectively; and
\textit{inconvenience} -- most sites merely provide the reference links to the source datasets and do not provide data utility tools like  covid19datahub~\cite{datahub} for easy access.

\subsection{Computational tool challenge}
There are existing resources that can assist users to identify malicious intent in websites. Google's Safe Browsing API, for instance, allows the user to enter a URL and check it against Google's constantly updated lists of unsafe web resources. Similar resources include isitPhishing.org, malwareurl.com, and antivirus software, among many others. Additionally, users can check malicious domain lists through different sources such as phishtank.com or the aforementioned Google's Safe Browsing lists. As many malicious sites use URL shorteners to disguise themselves, to counteract potential attacks, it would be safe to first use URL expanders to figure out what they are before clicking them. Despite the easy access of those computational tools, they are not available conveniently in a single place where different tools can be called up whenever needed. 

The awareness of these existing tools and efficient use of them for quick response is vital for combating COVID-19. An associate issue is the requirement for current and frequently updated black-lists~\cite{sahoo2017malicious}. As we know, it is infeasible to manually maintain a dynamically changing list of malicious URLs, with new sites being generated everyday. Therefore, it is necessary to develop AI/ML identifiers that can learn from the old malicious sites to estimate the threats of new ones.



\subsection{Ethics challenge}

The COVID-19 pandemic is ushering in a new era of digital surveillance since governments are employing tools that track and monitor individuals. South Korea and Israel, for instance, have demonstrated the effectiveness of harnessing different digital surveillance tools. However, such a new practice can breach data privacy in the meantime and may even remain in use after the pandemic. In this section, we discuss the potential \textit{privacy concerns}, \textit{trade-offs} between stringent disease monitoring and patient privacy and ethical issues behind the disruption of \textit{civil liberties}.

Gauging the war-like severity of the coronavirus pandemic, academics, researchers, companies and non-profits alike have come forward to contribute in any possible way. However, given the rapid nature of such responses and the subsequent lack of policy checks, these otherwise novel endeavors may have ethical loopholes. In an attempt to provide a transparent view of the degree of infection and prevent community spread of the virus, many counties and states in the United States have decided to publicly release data corresponding to cases, including the number of cases per zip-code~\cite{mallory2020sc}. Smartphone applications with geo-locating capabilities have come out for users to  log their symptoms. But the use of such applications has significant \textit{privacy concerns}~\cite{wetsman2020personal}. Contact tracing has been identified as an effective way to control the spread of the virus in communities where the infection is not yet widespread or has slowed down significantly, and companies including Google and Apple are currently developing applications to make this possible. Only when a sufficient number of people use the application and voluntarily report their cases can it be used as a reliable tool of tracking. In this situation, there is an obvious \textit{trade-off} between user health privacy and data transparency and it is challenging to identify well-defined ethical boundaries when it comes to public health during a pandemic. The success of such an app requires a majority of the population to download and use it.  


\section{Feasibility Discussion}

In this section, we present some current efforts that address the aforementioned challenges and show that the three challenges are solvable with collaborative research. 

\begin{figure}[tbph!]
    \centering
    \includegraphics[width=0.45\textwidth]{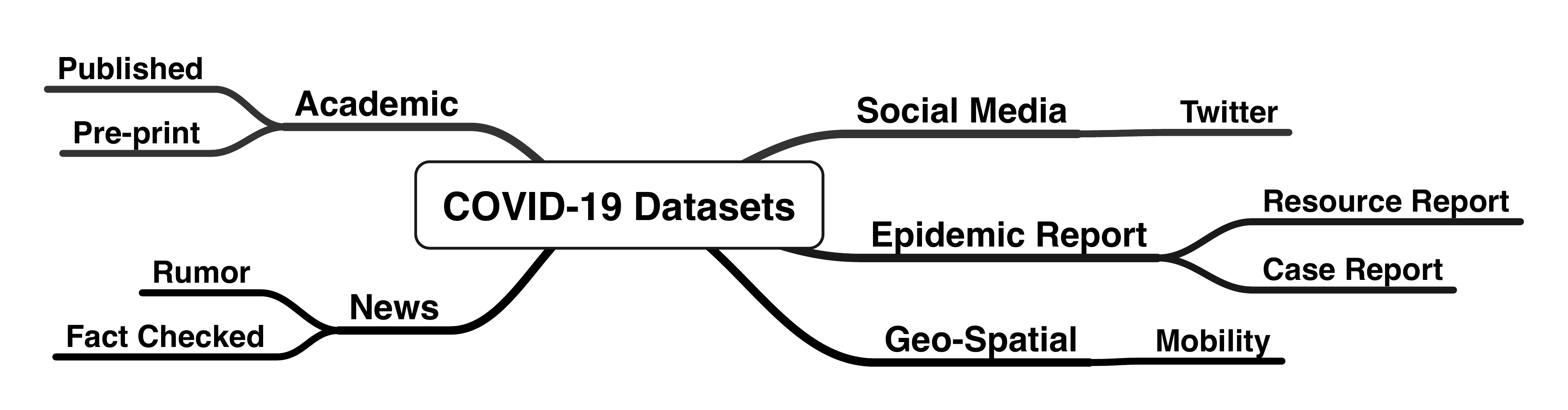}
    \caption{A taxonomy of collected datasets}
    \label{fig:tax}
\end{figure}

For the \textit{data challenge}, we collect the publicly available COVID-19 datasets and cluster them into several groups\footnote{ \href{https://github.com/bigheiniu/awesome-coronavirus19-dataset}{https://github.com/bigheiniu/awesome-coronavirus19-dataset}}. 
Under each group, researchers can reference complete datasets from different sources or settings. For example, in social media data, we gather available tweet corpus on COVID-19~\cite{b2020largescale}\cite{chen2020covid19} with different query keywords and time spans. The hierarchy cluster structure in Figure~\ref{fig:tax} helps the researchers to quickly locate the dataset. 
Lastly our data repository includes areas  in academics, news, social media, and epidemic reports for multi-disciplinary research. For example, if a researcher wants to analyze the influence of the news or academic findings on social media like Twitter, s/he can use the data in academic or news topics and social media. 
To help a researcher easily access the datasets in the repository, we build a data-loader\footnote{ \href{https://github.com/bigheiniu/COVID-19-Dataloaders}{https://github.com/bigheiniu/COVID-19-Dataloaders}}. It is a Python package with a pandas Dataframe~\cite{reback2020pandas} by calling \textit{data = DataLoader().download(url)}. This widely used data format can help the downstream data analysis. 

\begin{figure}[h!]
    \centering
    \includegraphics[width=0.45\textwidth]{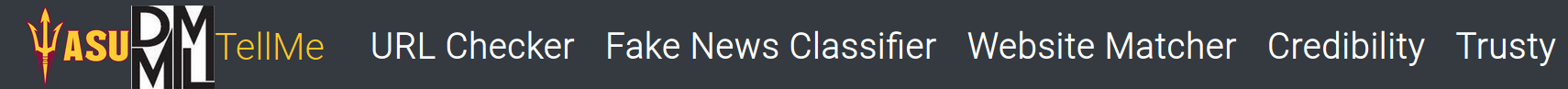}
    \caption{The components of the TellMe system.}
    \label{fig:TellMe}
\end{figure}

To tackle the too challenge, we develop \textit{TellMe}, a computational tool that provides an estimate if a piece of news or text is disinformation. Its input includes URLs and text, and its output is a score based on different functions of TellMe as shown in Figure~2: URL Checker, Fake News Classifier, Website Matcher, Credibility and Trusty. The Trusty~\cite{moturu2009quantifying} and Credibility~\cite{abbasi2013measuring} scores are based on contents' social engagements that malicious users share more similarity than general users. The fake news score is returned from a state-of-the-art fake new detector~\cite{shu2020leveraging}. The website matcher compares the input URL with websites that publish false information about the virus found by NewsGuard~\cite{brille2019newsguard}. 


Now, we use fake news as an example to illustrate our attempts to learn with weak social supervision to detect COVID-19 disinformation more effectively and with explainability. First, for \textit{effective fake news detection}, we consider the relationships among publishers, news pieces, and consumers, which is motivated by existing sociological studies on journalism on the correlation between the partisan bias of publishers, the credibility of consumers, and the veracity degree of news content; and explore various auxiliary information from these relations to help detect fake news~\cite{shu2019beyond}.  Second, for \textit{explainable fake news detection}, we aim to derive explanation of prediction results to help decision makers and practitioners; we attempt to explore user comments as a source and mine informative and relevant pieces to help explain why a piece of news is predicted as fake, and pinpoint more fictional text in news text simultaneously~\cite{shu2019defend}. 



To tackle the ethics challenge 
due to the increase in government surveillance and prevalence of smartphone apps to collect and gather user/patient data, we need to take into account legitimate concerns regarding privacy and the degree to which such a regime of monitoring and enforcement will affect democracy after the pandemic ends. It requires us to understand and acknowledge the fact that there is a clear difference between standard biomedical ethics versus privacy concerns and ethics during a public health crisis. Governments and public health officials may need to take certain measures aimed at minimizing the damage caused by the virus and for the common good during this trying time, which under normal circumstances might have been inappropriate. Nevertheless, measures could be taken to avoid potential misuse of data. One possible way to have better guarantees on user privacy would be to make these contact tracing smartphone applications communicate in an encrypted peer to peer way rather than storing all the data in a central server. These technologies should also be deployed in a way that is as transparent as possible, so that the user is fully aware of what and how much personal information he/she permits the application to use. Furthermore, there is significant ongoing discussion among experts, researchers and policy-makers regarding a steady recovery into a normal functioning society. For example, the ethics research group at Harvard University makes efforts at finding solutions without compromising user privacy to keep civil liberty and democracy at the forefront.

\section{Looking Ahead}

The significance of combating the COVID-19 infodemic lies at protecting people from falling victims to the pandemic in this unexpected front and from disrupting otherwise already inconvenient daily routines so as to improve our resilience in our fight to contain the pandemic. In this position paper, we show a good number of problems posed by the COVID-19 infodemic, the vast amounts of data generated in the world's effort to contain the pandemic, and the need for concerted efforts at various levels to efficiently and effectively deal with current and future challenges  in medical and information fronts.

It is evident that (1) we face both immediate and future challenges in this unprecedented fight, (2) existing data will grow fast, and existing computational tools are insufficient to contain and mitigate the COVID-19 infodemic, and (3) short-term solutions can have potential long-term impact. Therefore, when we face hard choices, we need to resist the temptation to trade-off so as to minimize long-term negative impact; when we search for solutions, we should consider those employing crowdsourcing and take long views for fairness and responsibility; when we design methods, we should rely on collective wisdom and diversity to aim for robustness; and when we form teams, we should give priority to multi-disciplinary collaboration and preemptively address hidden biases. Our future will always be uncertain, but with the advancement in science and technology and with our preparedness trained and tested in our concerted efforts to contain the pandemic in all fronts, our future will surely be brighter and healthier. 

\section*{Acknowledgments}
This work is, in part, supported by Global Security Initiative (GSI) at ASU and by NSF grants (\#2029044 and \#1614576). We would like to thank Denis Liu for helping develop earlier versions of TellMe  and for carefully proofreading an earlier version of this paper. 

\bibliographystyle{aaai}
\bibliography{aaai}

\end{document}